\documentclass[letterpaper,twocolumn,10pt]{article}
\usepackage{times}
\usepackage{graphicx}
\usepackage[hyphens]{url}
\PassOptionsToPackage{hyphens}{url}\usepackage{hyperref}
\usepackage{array}
\usepackage{setspace}
\usepackage[disable]{todonotes}
\usepackage{xspace}

\newcommand{\megele}[1]{\todo[inline]{megele: #1}}
\usepackage{pifont}
\usepackage{titlesec}
\titlespacing*{\subsection}{0pt}{.7\baselineskip}{.7\baselineskip}
\titlespacing*{\section}{0pt}{.7\baselineskip}{.7\baselineskip}

\newcommand{\totalNumProfiles}{7,250,228} 
\newcommand{\totalNumViolations}{966,306}
\newcommand{\totalNumDeleted}{400,389}

\newcommand{\compa}{{\sc Compa}\xspace}

\newcommand{\numTOCApps}{12,238}
\newcommand{\numApps}{302,513}

\newcommand{\totalTextOracleClusters}{374,920} 
\newcommand{\numTOCgood}{365,558} 
\newcommand{\numTOCbad}{9,362} 
\newcommand{\numTOCbadFP}{377} 
\newcommand{\numTOCbadFPR}{4} 
\newcommand{\numTOCbadAccounts}{343,229} 
\newcommand{\numTOCbadAccountsFP}{12,382}
\newcommand{\numTOCbadAccountsFPR}{3.6}
\newcommand{\numTOCbadApps}{257}

\newcommand{\numTOCBulk}{12,347} 
\newcommand{\numTOCBulkBad}{1,647} 
\newcommand{\numTOCBulkBadFP}{146} 
\newcommand{\numTOCBulkBadFPR}{8.9} 
\newcommand{\numTOCBulkBadAccounts}{178,557}
\newcommand{\numTOCBulkBadAccountsFP}{4,854}
\newcommand{\numTOCBulkBadAccountsFPR}{2.7}

\newcommand{\numTOCClient}{362,573} 
\newcommand{\numTOCClientBad}{7,715} 
\newcommand{\numTOCClientBadFP}{231} 
\newcommand{\numTOCClientBadFPR}{3.0} 
\newcommand{\numTOCClientBadAccounts}{164.672}
\newcommand{\numTOCClientBadAccountsFP}{7,528}
\newcommand{\numTOCClientBadAccountsFPR}{4.6}

\newcommand{\totalLinkOracleClusters}{14,548}

\newcommand{\numLOCbad}{1,236}
\newcommand{\numLOCbadFP}{72}
\newcommand{\numLOCbadFPR}{5.8}
\newcommand{\numLOCbadAccounts}{54,907}
\newcommand{\numLOCbadAccountsFP}{2,141}
\newcommand{\numLOCbadAccountsFPR}{3.8}

\newcommand{\numLOCBulk}{1,569}
\newcommand{\numLOCBulkBad}{251}
\newcommand{\numLOCBulkBadFP}{37}
\newcommand{\numLOCBulkBadFPR}{14.7}
\newcommand{\numLOCBulkBadAccounts}{8,254}
\newcommand{\numLOCBulkBadAccountsFP}{1,101}
\newcommand{\numLOCBulkBadAccountsFPR}{13.3}

\newcommand{\numLOCClient}{12,979}
\newcommand{\numLOCClientBad}{985}
\newcommand{\numLOCClientBadFP}{35}
\newcommand{\numLOCClientBadFPR}{3.5}
\newcommand{\numLOCClientBadAccounts}{46,653}
\newcommand{\numLOCClientBadAccountsFP}{1,040}
\newcommand{\numLOCClientBadAccountsFPR}{2.2}

\newcommand{\totalFacebookStati}{106,373,952}
\newcommand{\totalNumProfilesFB}{206,876}
 
\newcommand{\totalTextOracleClustersFB}{48,586}
\newcommand{\numTOCbadFB}{671}
\newcommand{\numTOCbadFPFB}{22}
\newcommand{\numTOCbadFPRFB}{3.3\%}
\newcommand{\numTOCbadAccountsFB}{11,499}
\newcommand{\numTOCbadAccountsFPFB}{412}
\newcommand{\numTOCbadAccountsFPRFB}{3.6\%}

\begin{document}

\title{Towards Detecting Compromised Accounts on Social Networks}
\author{Manuel Egele$^\dag$, Gianluca Stringhini$^\S$, Christopher Kruegel$^\ddag$, and Giovanni Vigna$^\ddag$\\
\ \\
$^\dag$Boston University\hfill
$^\S$University College London\hfill
$^\ddag$UC Santa Barbara\\
\textit{megele@bu.edu}, \textit{g.stringhini@ucl.ac.uk}, \textit{\{chris,vigna\}@cs.ucsb.edu}
}
\date{}

\maketitle
\begin{abstract}
  Compromising social network accounts has become a profitable course of action
  for cybercriminals. By hijacking control of a popular media or business
  account, attackers can distribute their malicious messages or disseminate fake
  information to a large user base. The impacts of these incidents range from a
  tarnished reputation to multi-billion dollar monetary losses on financial
  markets.
  In our previous work, we demonstrated how we can detect large-scale compromises
  (i.e., so-called campaigns) of regular online social network users. In this
  work, we show how we can use similar techniques to identify compromises of
  individual high-profile accounts. High-profile accounts frequently have one
  characteristic that makes this detection reliable -- they show consistent
  behavior over time.
  We show that our system, were it deployed, would have been able to detect and
  prevent three real-world attacks against popular companies and news agencies.
  Furthermore, our system, in contrast to popular media, would not have fallen
  for a staged compromise instigated by a US restaurant chain for publicity
  reasons.
\end{abstract}

\thispagestyle{empty}
\pagestyle{empty}
\section{Introduction}
Online social networks, such as Facebook and Twitter, have become one of the
main media to stay in touch with the rest
of the world. Celebrities use them to communicate with their fan base,
corporations take advantage of them to promote their brands and have a
direct connection to their customers, while news agencies leverage social
networks to distribute breaking news. Regular users make pervasive use of
social networks too, to stay in touch with their friends or colleagues and
share content that they find interesting. 

Over time, social network users build trust relationships with the accounts they
follow.
This trust can develop for a variety of reasons.
For example, the user might know the owner of the trusted account in person or
the account might be operated by an entity commonly considered as trustworthy, such as
a popular news agency.
Unfortunately, should the control over an account fall into the hands of a cyber
criminal, he can easily exploit this trust to further his own malicious agenda.
Previous research showed that using compromised accounts to spread malicious
content is advantageous to cyber criminals, because social network users are
more likely to react to messages coming from accounts they
trust~\cite{phishing}.

These favorable probabilities of success exceedingly attract the attention of
cyber criminals.
Once an attacker compromises a social network account he can use it for
nefarious purposes such as sending spam messages or link to malware and
phishing web sites~\cite{Grier:2010:ccs}.
Such traditional attacks are best carried out through a large population of
compromised accounts belonging to regular social network account users.
Recent incidents, however, demonstrate that attackers can cause havoc and
interference even by compromising individual, but high-profile accounts.
These accounts (e.g., newspaper or popular brand name accounts) have large social
circles (i.e., followers) and their popularity suggests trustworthiness to many
social network users.
Recent attacks show that compromising these high profile accounts can be
leveraged to disseminate fake news alerts, or messages that tarnish a company's
reputation~\cite{
foxnewspolitics:compromise,ap_hack,theonion:compromise,skype:compromise}.

Moreover, the effects of an account compromise can extend well beyond the
reputation of a company.
For example, the dissemination of an erroneous Associated Press (AP) news story
about a bomb exploding in the White House in 2013 led to a 1\% drop in the
Standard \& Poor's 500 index, temporarily wiping out US\$ 136B~\cite{bloomberg_ap_hack}.
Compromises of high profile accounts usually get cleaned up quickly after they
are detected.
Unfortunately, since detection is still exclusively a manual endeavor, this is
often too late to mitigate the negative impacts of account compromises.
For example, the above mentioned AP message was shared by over 3,000 users
before the compromise was detected and the offending message removed.
Similarly, a message sent as a consequence of a compromise of the Skype Twitter
account happening during a national holiday remained accessible for over a
day~\cite{skype:compromise}.
These incidents show that it is critical for a social network to be able to
reliably detect and block messages that have not been authored by an account's
legitimate owner.

A wealth of research was proposed in the last years to detect malicious 
activity on online social networks. Most of these systems, however, focus on 
detecting fake accounts specifically created to spread malicious content, 
instead of looking for legitimate accounts that have been compromised~\cite{
Benvenuto:2010:CEAS,Lee:2010:SIGIR,Stringhini:10:socialnet-spam}.
These systems are inadequate to detect compromised accounts, because legitimate,
yet compromised 
accounts have significantly different characteristics than fake ones.
Other mitigation techniques have a more general scope, and either detect 
malicious accounts by grouping together similar messages~\cite{Gao:2012:ndss,
Gao:2010:imc} or by looking at the presence of suspicious URLs in 
social network messages~\cite{Lee:2012:ndss,Thomas:2011:oakland}. These 
systems can detect messages that are sent by compromised social network 
accounts, in case cybercriminals use multiple accounts to send similar 
messages, or the messages are used to advertise web pages pointing to malware 
or phishing. In the case of the high-profile compromises mentioned before, 
however, neither of these conditions apply: the compromises each consisted of a 
single message, and no URLs were contained in any of the messages. Therefore, 
previously-proposed systems are inadequate to detect this type of compromises.

In this paper we present \compa{}, the first detection system designed to
identify compromised social network accounts. \compa{} is based on a simple
observation: social network users develop habits over time, and these habits are
fairly stable. A typical social network user, for example, might consistently
check her posts in the morning from her phone, and during the lunch break from
her desktop computer. Furthermore, interaction will likely be limited to a
moderate number of social network contacts (i.e., friends).  Conversely, if the
account falls under the control of an adversary, the messages that the attacker
sends will likely show anomalies compared to the typical behavior of
the user. 

To detect account compromises, \compa{} builds a behavioral profile for social
network accounts, based on the messages sent by the account in the past.  Every
time a new message is generated, the message is compared against this behavioral
profile. If the message significantly deviates from the learned behavioral
profile, \compa{} flags it as a possible compromise.  
In this paper we first show that high profile accounts often have well-defined
behavioral profiles that allow \compa to detect compromises with very low false
positives.
However, behavioral profiles of regular user accounts are more variable than
their well-defined counterparts of most high profile accounts.
This is because regular users are more likely to experiment with new features or
client software to engage with the social network.
This variability could cause an increase of false positive alerts.
However, social network accounts of regular users are less influential than high
profile accounts.
Thus, attackers aggregate multiple accounts into a {\em campaign} to achieve
effects that are similar to the compromise of a high profile account.
\compa uses this insight to first identify campaigns by means of message
similarity and only labels accounts as compromised if a significant portion of
messages in a campaign violate the behavioral profile of their underlying
account.
This allows us to keep false positives low, while still being able to detect
accounts that are victims of large-scale compromises.

To evaluate \compa{}, we applied it to four Twitter compromises that affected
high profile accounts over the last three years.
We show that our system would have been able to detect those malicious messages
before they were posted, avoiding the fake information to spread.
We also show a case study of a compromise that was faked by the Chipotle Twitter
account for promotional reasons; in this case \compa{} correctly detected that
the alleged malicious messages did not deviate from the regular behavior of the
account.
Finally, we also applied \compa{} to two datasets from Twitter and Facebook,
looking for large-scale compromises.
The Twitter dataset consists of 1.4 billion messages we collected from May 13,
2011 to August 12, 2011, while the Facebook dataset contains 106 million
messages ranging from September 2007 to July 2009 collected from several large
geographic networks.
Our results show that \compa{} is effective in detecting compromised accounts
with very few false positives.
In particular, we detected 383,613 compromised accounts on Twitter, and 11,087
compromised accounts on Facebook.

In summary, this paper makes the following contributions:
\begin{itemize}
 \item We present \compa{}, the first system designed to detect compromised
   social network accounts.
 \item We show that \compa{} can reliably detect compromises that affect high
   profile accounts. Since the behavior of these accounts is very consistent,
   false positives are minimal.
 \item To detect large-scale compromises, we propose to group similar messages
   together and apply \compa{} to them, to assess how many of those messages
   violate their accounts' behavioral profile. This grouping accounts for
   the fact that regular social network accounts show a more variable behavior
   compared to high profile ones, and allows us to keep false positives low.
\item We apply \compa{} to two datasets from popular social networks, Facebook
  and Twitter, and show that our system would have been able to detect hundreds
  of thousands of compromised accounts. We also show that \compa would have been 
able to detect four high-profile compromises that affected popular Twitter accounts, and 
to correctly flag as legitimate a fake compromise that was attempted by a US fast food 
chain on their Twitter account for promotional reasons. 
\end{itemize}

\noindent\textbf{Comparison with previous published version}. This paper is the extended version of our previous work~\cite{egele13:compa} 
that was published at the Network and Distributed Systems Security Symposium 
in 2013. Compared to the original paper, in which we focused on large-scale 
compromises that affect thousands of social network accounts at the same 
time, in this paper we also look at isolated compromises that affect high-
profile accounts. We show that such accounts typically show a very consistent 
behavior, and therefore \compa{} can reliably detect compromises against 
them. To demonstrate this, we analyzed four compromises of high-profile 
accounts that made the news during the past three years, showing that \compa 
would have detected them.

\section{Background: Social Network Compromises} 
\label{sec:background} 

In the following, we illustrate four case studies where high-profile Twitter
accounts were compromised.
We will use these case studies to both show how critical a social network
compromise can be for a company, as well as how our system could be used to
detect and ultimately prevent such attacks.

\noindent\textbf{Associated Press.} On April $23^{rd}$ 2013, the Twitter account of the Associated Press (@AP) was
compromised~\cite{ap_hack}.
The account was misused to distribute false information about president Obama
being hurt by an explosion in the White House.
This message had an interesting side effect: seconds after being posted, it was
used as a signal of negative events by automated trading bots on the New York
stock exchange.
This signal lead to a perceivable drop in the market index which recovered after
the information was confirmed to be false~\cite{bloomberg_ap_hack}.
This incident shows how a social network compromise can have significant effects
on the real world.

\noindent\textbf{FoxNews Politics.} On July $4^{th}$ 2011, the Twitter account
of Fox News' politics (@foxnewspolitics) division got
compromised~\cite{foxnewspolitics:compromise}. The attackers used this
opportunity to distribute the information that president Obama got assassinated.

\noindent\textbf{Skype.}
On new year's day 2014, the Twitter account of the Skype Voip service was
compromised.
The attacker used his access to discourage the use of Mircrosoft's email
products for the fear of disclosing information to government agencies.
We would assume that an observant legitimate owner of the account would
detect such a malicious message during their regular activity.
However, presumably because of the holiday season, it took more than two hours
before the offending message was removed by the legitimate owners of the Skype
account.
In the meantime, the offending message got retweeted over 8,000 times.
This incident prominently demonstrates the advantages an automated technique for
the detection of compromised accounts would entail, as such attacks can have
significant negative impact on a brand's online reputation. 

\noindent\textbf{Yahoo! News.} More recently, in August 2014, Yahoo!'s news
account (@YahooNews) also got compromised and used to disseminate false
information regarding an Ebola outbreak in Atlanta, GA. 

To prevent social network accounts from being compromised, we propose to learn the 
typical behavior of a user, and flag a message as a possible compromise if it does not 
match the learned behavior. In the following section, we describe in detail the 
behavioral profile that we leverage as part of our system. In Section~\ref{sec:casestudies} 
we provide details on the anomalies generated by the four described 
high-profile incidents, which allowed \compa to detect them.

\section{Behavioral Profiles}
\label{sec:profiles}

A behavioral profile 
leverages historical information
about the activities of a social network user to capture this user's
normal (expected) behavior. To build behavioral profiles, our system
focuses on the stream of messages that a user has posted on the social
network.  Of course, other features such as profile pictures or social activity
(e.g., establishing friend or follower relationships) could be useful 
as well. Unfortunately, social networks
typically do not offer a way to retrieve historical data about changes
in these features, and therefore, we were unable to use them.

A behavioral profile for a user $U$ is built in the following way:
Initially, our system obtains the stream of messages of $U$ from the social
networking site. The message stream is a list of all
messages that the user has posted on the
social network, in chronological order. For different social networks, the message streams are
collected in slightly different ways. For example, on Twitter, the
message stream corresponds to a user's public timeline. For Facebook,
the message stream contains the posts a user wrote on her own wall, but it also
includes the messages that this user has posted on her friends' walls.

To be able to build a comprehensive profile, the stream needs to
contain a minimum amount of messages. Intuitively, a good behavioral
profile has to capture the breadth and variety of ways in which a
person uses her social network account (e.g., different client applications or languages). Otherwise, an incomplete
profile might incorrectly classify legitimate user activity as
anomalous. 
Therefore, we do not create behavioral
profiles for accounts whose stream consists of less than a minimum
number {\em S} of messages. In our experiments, we empirically
determined that a stream consisting of less than $S=10$ messages
does usually not contain enough variety to build a representative
behavioral profile for the corresponding account. Furthermore, profiles that
contain less then {\em S} messages pose a limited threat to the social network
or its users. This is because such accounts are either new or very inactive and
thus, their contribution to large scale campaigns is limited. A detailed discussion of
this threshold is provided in our previous work~\cite{egele13:compa}.

Once our system has obtained the message stream for a user, we use
this information to build the corresponding behavioral profile. More
precisely, the system extracts a set of feature values from each
message, and then, for each feature, trains a statistical model. Each
of these models captures a characteristic feature of a message, such
as the time the message was sent, or the application that was used to
generate it. The features used by these models, as well as the models
themselves, are described later in this section.

Given the behavioral profile for a user, we can assess to what extent
a new message corresponds to the expected behavior. To this end, we
compute the anomaly score for a message with regard to the user's established
profile. The anomaly score is computed by extracting the feature
values for the new message, and then comparing these feature values to the corresponding
feature models. Each model produces a score (real value) in the
interval $[0,1]$, where $0$ denotes perfectly normal (for the feature
under consideration) and 1 indicates that the feature is highly
anomalous. The anomaly score for a message is then calculated by
composing the results for all individual models.

\subsection{Modelling Message Characteristics}
\label{sec:models}

Our approach models the following seven features when building a
behavioral profile. 

\smallskip \noindent {\bf Time (hour of day).}
This model captures the hour(s) of the day during which an account is
typically active. Many users have certain periods during the course of
a day where they are more likely to post (e.g., lunch breaks) and others that are typically
quiet (e.g., regular sleeping hours). 
If a user's stream indicates regularities in
social network usage, messages that appear during hours that are
associated with quiet periods are considered anomalous.

\smallskip \noindent {\bf Message Source.}
The source of a message is the name of the application that was used
to submit it. Most social networking sites offer traditional web and
mobile web access to their users, along with applications for mobile platforms such as iOS and Android.
Many social network ecosystems provide access to a 
multitude of applications
created by independent, third-party developers. 

Of course, by default, a third-party application cannot post messages
to a user's account. However, if a user chooses to, she can grant this
privilege to an application. The state-of-the-art method of governing
the access of third-party applications is 
 \textsc{OAuth}~\cite{oauth}. \textsc{OAuth} is implemented by Facebook and
Twitter, as well as numerous other, high-profile web sites, and enables a user
to grant access to her profile without revealing her credentials.

By requiring all third-party
applications to implement \textsc{OAuth}, the social network operators can easily
shut down individual applications, should that become necessary. In
fact, our evaluation shows that third-party applications are
frequently used to send malicious messages. 

This model determines whether a user has
previously posted with a particular application or whether this is the
first time. Whenever a user posts a message from a new application,
this is a change that could indicate that an attacker has succeeded
to lure a victim into granting access to a malicious application.

\smallskip \noindent {\bf Message Text (Language).}
A user is free to author her messages in any language. However, we
would expect that each user only writes messages in a few languages
(typically, one or two).  Thus, especially for profiles where this
feature is relatively stable, a change in the language 
is an indication of a suspicious change in user activity.

To determine the language that a message was written in, we leverage
the {\tt libtextcat} library. This library performs n-gram-based text
categorization, as proposed by Cavnar and Trenkle~\cite{cavnar}.
Of course, for very short messages, it is often difficult to determine
the language. This is particularly problematic for Twitter messages,
which are limited to at most 140 characters and frequently contain abbreviated
words or uncommon spelling.

\smallskip \noindent {\bf Message Topic.}
Users post many messages that contain chatter or mundane information. 
But we would also expect that many users
have a set of topics that they frequently talk about, such as favorite
sports teams, music bands, or TV shows. When users typically focus on
a few topics in their messages and then suddenly post about some
different and unrelated subject, this new message should be rated as
anomalous. 

In general, inferring message topics from short snippets of text without context
is difficult. However, some social networking platforms allow users to label messages to explicitly
specify the topics their messages are about. When such labels or tags
are available, they provide a valuable source of information. A
well-known example of a message-tagging mechanism are Twitter's
\emph{hashtags}. By prefixing the topic keyword with a hash character a user
would use \#Olympics to associate her tweet with the Olympic Games. Using
hashtags to identify topics in messages have become so popular that Facebook
decided in August 2013 to incorporate this feature unmodified.

More sophisticated (natural language processing) techniques to extract message
topics are possible. However, such techniques are out of scope of this 
work.

\smallskip \noindent {\bf Links in Messages.}
Often, messages posted on social networking sites contain links to
additional resources, such as blogs, pictures, videos, or news
articles. Links in messages of social networks are so common that
some previous work has strongly focused on the analysis of URLs, often as
the sole factor, to determine whether a message is malicious or
not. We also make use of links as part of the behavioral profile of a
user. However, in our system the link information only represents a single
dimension (i.e., feature) in the feature vector describing a message. Moreover, recall that our features are primarily concerned with capturing the
normal activity of users. That is, we do not attempt to detect whether a
URL is malicious in itself but rather whether a link is different than
what we would expect for a certain user.

To model the use of links in messages, we only make use of the domain
name in the URL of links. The reason is that a user might regularly
refer to content on the same domain. For example, many users tend to
read specific news sites and blogs, and frequently link to interesting
articles there. Similarly, users might have preferences for a certain URL
shortening service. Of course, the full link differs among these
messages (as the URL path and URL parameters address different,
individual pages). The domain part, however, remains constant. Malicious
links, on the other hand, point to sites that have no legitimate use.
Thus, messages that link to domains that have not been observed in the
past indicate a change. The model also considers the general frequency
of messages with links, and the consistency with which a user links to
particular sites.

\smallskip \noindent {\bf Direct User Interaction.}
Social networks offer mechanisms to directly interact
with an individual user. The most common way of doing this is by
sending a direct message that is addressed to the recipient. Different
social networks have different mechanisms for doing that. For example,
on Facebook, one posts on the recipient user's wall; on Twitter, it is
possible to directly ``mention'' other users by putting the {\em @}
character before the recipient's user name. Over time, a user builds a
personal interaction history with other users on the
social network. This feature aims to capture the interaction
history for a user. In fact, it keeps track of the users an account ever
interacted with. Direct messages are sent to catch the attention of their
recipients, and thus are frequently used by spammers.

\smallskip \noindent {\bf Proximity.}
In many cases, social network users 
befriend other users that are geographically or contextually 
close to them. For example, a typical Facebook user will have many friends
that live in the same city, went to the same school, or work for the same
company. If this user suddenly started interacting with people who live on
another continent, this could be suspicious.
Some social networking sites (such as Facebook) express this proximity notion by grouping their
users into networks. The proximity model looks at the messages sent by a user. If a user sends a
message to somebody in the same network, this message is considered as local.
Otherwise, it is considered as not local. 
This feature captures the fraction of local vs. non-local messages.

\section{Training and Evaluation of the Models}
\label{sec:training}

In a nutshell, \compa{}
works as follows: for each social network user, we retrieve the past messages that the 
user has authored. We then extract features for each message, and build behavioral 
models for each feature separately. Then, we assess whether each individual feature 
is anomalous or not, based on previous observations. Finally, we combine the anomaly 
scores for each feature to obtain a global anomaly score for each message. This
score indicates whether the account has likely been 
compromised. In the following, we describe our approach in more detail.

\smallskip \noindent {\bf Training.} The input for the training step
of a feature model is the series of messages (the message stream) that were
extracted from a user account. For each message, we extract the
relevant features such as the source application and the domains of all links. 

Each feature model is represented as a set $\textbf{M}_f$. Each element
of $\textbf{M}_f$ is a tuple $<{fv}, c>$. ${fv}$ is the value of a
feature (e.g., {\tt English} for the language model, or
{\tt example.com} for the link model). $c$ denotes the number of
messages in which the specific feature value ${fv}$ was present. In
addition, each model stores the total number $N$ of messages that were
used for training.

\smallskip \noindent Our models fall into two categories:

\begin{itemize}
\item \emph{Mandatory} models are those where there is one feature
  value for each message, and this feature value is always present. 
  Mandatory models are \emph{time of the day}, \emph{source},
  \emph{proximity}, and \emph{language}.

\item \emph{Optional} models are those for which not every message has to have a
  value. Also, unlike for mandatory models, it is possible that
  there are multiple feature values for a single message. Optional models are
  \emph{links}, \emph{direct interaction}, and \emph{topic}. For
  example, it is possible that a message contains zero, one, or multiple
  links. For each optional model, we reserve a specific element with
  ${fv}$ = {\tt null}, and associate with this feature value the
  number of messages for which no feature value is present (e.g., the
  number of messages that contain no links).

\end{itemize}

The training phase for the \emph{time of the day} model works slightly
differently. Based on the previous description, our system would first
extract the hour of the day for each message. Then, it would store,
for each hour $fv$, the number of messages that were posted during
this hour. This approach has the problem that strict one hour intervals, unlike the
progression of time, are discrete. Therefore, messages that are sent
close to a user's ``normal'' hours could be incorrectly considered as
anomalous.

To avoid this problem, we perform an adjustment step after the
\emph{time of the day} model was trained (as described above). In
particular, for each hour $i$, we consider the values for the two adjacent
hours as well. That is, for each element $<i, c_i>$ of $\textbf{M}_f$, a
new count $c\prime_i$ is calculated as the average between the number
of messages observed during the $i^{th}$ hour ($c_i$), the number of
messages sent during the previous hour ($c_{i-1}$), and the ones
observed during the following hour ($c_{i+1}$). After we computed all
$c\prime_i$, we replace the corresponding, original values
in $\textbf{M}_f$.

As we mentioned previously, we cannot reliably build a behavioral profile if
the message stream of a user is too short. Therefore, the training phase is
aborted for streams shorter than $S=10$, and any message sent by those users is not
evaluated.

\smallskip \noindent {\bf Evaluating a new message.} 
When calculating the anomaly score for a new message, we want to evaluate whether this message violates the behavioral
profile of a user for a given model. In general, a message is considered more anomalous
if the value for a particular feature did not appear at all in the stream of a
user, or it appeared only a small number of times.
For {\it mandatory features}, the anomaly score of a message is calculated
as follows:

\begin{enumerate}
\item The feature $fv$ for the analyzed model is first extracted from the message.
If $\textbf{M}_f$ contains a tuple with $fv$ as a first element, then the tuple $<fv, c>$ is extracted from
$\textbf{M}_f$. If there is no tuple in $\textbf{M}_f$ with $fv$ as a first value,
the message is considered anomalous. The procedure terminates here and
an anomaly score of 1 is returned. 
\item As a second step, the approach checks if $fv$ is anomalous at all for the
behavioral profile built for the feature under consideration. $c$ is compared
to $\bar{M_f}$, which is defined as
$\bar{M_f} = \frac{\sum_{i=1}^{\| \textbf{M}_f \|}c_i}{N}$,
where $c_i$ is, for each tuple in $\textbf{M}_f$, the second element of the tuple.
If $c$ is greater or equal than $\bar{M_f}$, the message is considered to
comply with the learned behavioral profile for that feature, and an anomaly score
of 0 is returned. The rationale behind this is that, in the past, the user has shown a
significant number of messages with that particular $fv$.
\item If $c$ is less than $\bar{M_f}$, the message is considered somewhat
anomalous with respect to that model. Our approach calculates the relative
frequency $f$ of $fv$ 
as
$f = \frac{c_{fv}}{N}$.
The system returns an anomaly score of 1 - $f$.
\end{enumerate}

\smallskip \noindent The anomaly score for {\it optional features} is calculated
as:

\begin{enumerate}
\item The value $fv$ for the analyzed feature is first extracted from the message.
If $\textbf{M}_f$ contains a tuple with $fv$ as a first element, the message is 
considered to match the behavioral profile, and an anomaly score of 0 is returned.
\item If there is no tuple in $\textbf{M}_f$ with $fv$ as a first element,
the message is considered anomalous.
The anomaly score in this case is defined as the probability $p$ for the account
to have a {\tt null} value for this model. Intuitively, if a user rarely uses a
feature on a social network, a message containing an $fv$ that has never been
seen before for this feature is highly anomalous.
The probability $p$ is calculated as 
 $p = \frac{c_{null}}{N}$.
If $\textbf{M}_f$ does not have a tuple with {\tt null} as a first
element, $c_{null}$ is considered to be 0. $p$ is then
returned as the anomaly score.
\end{enumerate}

As an example, consider the following check against the \emph{language} model:
The stream of a particular user is composed of 21 messages. 
Twelve of them are in English, while nine are in German.
The $\textbf{M}_f$ of the user for that particular model looks like this:
\begin{center}
($<$English,12$>$,$<$German,9$>$).
\end{center}
The next message sent by that user will match one of three cases:
\begin{itemize}
\item The new message is in English. Our approach extracts the tuple
$<$English,12$>$ from $\textbf{M}_f$, and compares $c = 12$ to $\bar{M} = 10.5$.
Since $c$ is greater than $\bar{M_f}$, the message is considered normal, and an
anomaly score of 0 is returned.
\item The new message is in Russian. Since the user never sent a message in that
language before, the message is considered very suspicious, and an anomaly score of 1
is returned.
\item The new message is in German. Our approach extracts the tuple $<$German,
9$>$ from $\textbf{M}_f$, and compares $c = 9$ to $\bar{M_f} = 10.5$. Since $c <
\bar{M_f}$, the message is considered slightly suspicious. The
relative frequency of German tweets for the user is 
$f = \frac{c}{N} = 0.42$.
Thus, an anomaly score of $1 - f = 0.58$ is returned. This means that the
message shows a slight anomaly in the user average behavior. However, as 
explained in Section~\ref{sec:detection}, on its own this score will not be enough to flag the
message as malicious. 
\end{itemize}

\smallskip \noindent {\bf Computing the final anomaly score.}  Once
our system has evaluated a message against each individual feature model, we
need to combine the results into an overall anomaly score for this message. This
anomaly score is a weighted sum of the values for all models. We use
Sequential Minimal Optimization~\cite{smo} to learn the optimal
weights for each model, based on a training set of instances (messages and
corresponding user histories) that are labeled as malicious and
benign.  Of course, different social networks will require different
weights for the various features. A message is said to violate an
account's behavioral profile if its overall anomaly score exceeds a
threshold.  In Section~\ref{sec:labels}, we present a more detailed
discussion on how the features and the threshold values were calculated. More
details, including a parameter sensitivity analysis on the threshold value, are
presented in our previous work~\cite{stringhini2013follow,poultry}. Moreover, we
discuss the weights (and importance) of the features for the
different social networks that we analyzed (i.e., Twitter and Facebook).

\smallskip \noindent{\bf Robustness of the Models.}
In our original paper we show that it is difficult for an attacker to mimic all the behavioral models used by \compa~\cite{egele13:compa}.
In addition, in our setup we only used features that are observable from the outside --- if \compa was deployed by a social network instead, they could use additional indicators, such as the IP address that a user is connecting from or the browser user agent.
\smallskip \noindent{\bf Novelty of the modelled features.}
\label{sec:novelty}
In our previous paper~\cite{egele13:compa} we show that most of the features 
used by \compa are novel, and were not used by previous work. In addition, 
existing systems focus on detecting fake accounts, and therefore look for 
similarities across different accounts to flag them as malicious. In \compa, 
conversely, we look for changes in the behavior of legitimate accounts.

\subsection{Training the Classifier} \label{sec:labels}
As discussed in Section~\ref{sec:training}, \compa uses a weighted sum of
feature values to determine whether a new message violates the behavioral
profile of its social network account.
Naturally, this bears the question how to determine optimal feature weights to
calculate the weighted sum itself.
To determine the feature weights in \compa, we applied Weka's SMO~\cite{weka} to
a labeled training dataset for both Twitter and Facebook.
A detailed discussion how we prepared the training datasets can be found in our
previous work~\cite{egele13:compa}. Note that this dataset is different than the one used to evaluate \compa in Section~\ref{sec:evaluation}.

While on Facebook, at the time of our experiment, we could easily infer a user location from her geographic networks, 
Twitter does not provide such a convenient proximity feature. Therefore, we
omitted this feature from the evaluation on Twitter. For Twitter, the
weights for the features are determined from a labeled training dataset
consisting of 5,236 (5142 legitimate, 94 malicious) messages with their
associated feature values as follows: {\em Source (3.3)}, {\em Personal
  Interaction (1.4)}, {\em Domain (0.96)}, {\em Hour of Day (0.88)},
{\em Language (0.58)}, and {\em Topic (0.39)}.

On Facebook, based on a labeled training dataset of 279 messages 
  (181 legitimate, 122 malicious), the weights were: {\em Source
  (2.2)}, {\em Domain (1.1)}, {\em Personal Interaction (0.13)}, {\em
  Proximity (0.08)}, and {\em Hour of Day (0.06)}. Weka determined
that the {\em Language} feature has no effect on the classification.
Moreover, as discussed earlier, assessing the message topic of an
unstructured message is a complicated natural language processing
problem. Therefore, we omitted this feature from the evaluation on the
Facebook dataset. 

\section{Behavioral Profile Stability}
\label{sec:profile_stability}

Detecting deviations in account behavior is simplified if the commonly occurring
behavior follows mostly regular patterns.
Thus, in this section we ask (and answer) the question of whether there is a
class of social network accounts that are particularly amenable to such an
analysis.
Arguably, a social network strategy is a crucial part for the public relation
department of most contemporary companies.
Intuitively, we would expect a well managed company account to show a more
stable behavior over time than accounts operated by regular users.
To assess whether this intuition is valid we conducted an experiment and
evaluated the message streams of popular companies for behavioral profile
violations. As positive example of social network compromises, 
we considered the four high-profile incidents described previously. 
As a baseline comparison we also evaluated the message streams of randomly
chosen social network accounts. 

\begin{table}[t]
\centering
\scalebox{0.6}{
\begin{tabular}{|c|l|r|c|l|r|}
\hline
\# & Twitter Account & Violations (\%) & \# & Twitter Account & Violations (\%) \\
\hline
1 & 163 & 0\% & 40 & derspiegel & 2\% \\
2 & alibabatalk & 0\% & 41 & espn & 2\% \\
3 & ap & 0\% & 42 & imgur & 2\% \\
4 & bloombergnews & 0\% & 43 & msnbc & 2\% \\
5 & bostonglobe & 0\% & 44 & tripadvisor & 2\% \\
6 & bw & 0\% & 45 & twitch & 2\% \\
7 & ebay & 0\% & 46 & xe & 2\% \\
8 & ehow & 0\% & 47 & yahoosports & 2\% \\
9 & engadget & 0\% & 48 & walmart & 2\% \\
10 & expedia & 0\% & 49 & bing & 3\% \\
11 & forbes & 0\% & 50 & nfl & 3\% \\
12 & foxnews & 0\% & 51 & reverso\_ & 3\% \\
13 & foxnewspolitics & 0\% & 52 & blizzardcs & 4\% \\
14 & gsmarena\_com & 0\% & 53 & google & 4\% \\
15 & huffingtonpost & 0\% & 54 & linkedin & 4\% \\
16 & imdb & 0\% & 55 & yahoofinance & 4\% \\
17 & latimes & 0\% & 56 & cnn & 5\% \\
18 & lemondefr & 0\% & 57 & timeanddate & 5\% \\
19 & msn & 0\% & 58 & yandexcom & 5\% \\
20 & nbcnews & 0\% & 59 & urbandictionary & 5\% \\
21 & nytimes & 0\% & 60 & netflix & 6\% \\
22 & pchgames & 0\% & 61 & weebly & 6\% \\
23 & reuters & 0\% & 62 & stumbleupon & 7\% \\
24 & skype & 0\% & 63 & yahooanswers & 7\% \\
25 & stackfeed & 0\% & 64 & reddit & 9\% \\
26 & steam\_games & 0\% & 65 & yelp & 9\% \\
27 & washingtonpost & 0\% & 66 & instagram & 10\% \\
28 & yahoo & 0\% & 67 & youtube & 10\% \\
29 & 9gag & 1\% & 68 & nih & 12\% \\
30 & amazon & 1\% & 69 & ancestry & 13\% \\
31 & digg & 1\% & 70 & microsoft & 13\% \\
32 & el\_pais & 1\% & 71 & paypal & 13\% \\
33 & facebook & 1\% & 72 & tumblr & 15\% \\
34 & ign & 1\% & 73 & wikipedia & 15\% \\
35 & internetarchive & 1\% & 74 & wordpress & 28\% \\
36 & pinterest & 1\% & 75 & AskDotCom & 39\% \\
37 & yahoonews & 1\% & 76 & bookingcom & 44\% \\
38 & abcnews & 2\% & 77 & twitter & 46\% \\
39 & bbcnews & 2\% & 78 & guardian & 47\% \\
\hline
\end{tabular}
}
\caption{\small{Behavioral profile violations of news agency and corporate Twitter accounts within
most recent 100 tweets.}}
\label{tbl:consistency_behavior}
\end{table}


\subsection{Popular Accounts} \label{sec:popular_accounts}

To assess whether the behavioral profiles of popular accounts are indeed mostly
stable over time we performed the following experiment.
Alexa~\cite{alexa} is a service that ranks popular websites.
We assume that most popular websites are operated by popular businesses.
Thus we identify the Twitter accounts that correspond to the top 5 entries in
each of 16 categories ranked by Alexa (e.g., arts, news, science, etc.).
Additionally, we add the Twitter accounts that correspond to the top 50 entries
of Alexa's top 500 global sites.
While a more exhaustive list would be beneficial, identifying a social network
account that corresponds to a website is a manual process and thus does not
scale well.
Table~\ref{tbl:consistency_behavior} presents the list of the resulting 78
Twitter accounts after removal of duplicate entries cross listed in multiple
categories.

For each account in this list \compa then built the behavioral profile and
compared the most recent 100 messages against the extracted profile.
As for any detection system, \compa needs to make tradeoffs between false positives and false negatives. To tune our system,
we used as ground truth the 4 high-profile incidents described in Section~\ref{sec:background}. We configured \compa to detect such attacks. 
We then analyzed the false positive rate 
that \compa generates by using this threshold. Note that since these incidents are the 
only ones that have been reported for the involved  accounts, this experiment
resulted in no false negatives.

Table~\ref{tbl:consistency_behavior} also shows how many of these 100 messages
violated their behavioral profile.
The results indicate that the majority of popular accounts have little
variability in their behavior.
As we can see, the majority of the high profile accounts that we 
evaluated have a very consistent behavior. In fact, as we will show in the 
next section, such accounts show a considerably more consistent behavior than 
average social network accounts. In these cases \compa could protect these 
accounts and still reliably detect compromises without fearing false positives.

A handful of high profile accounts, however, showed a very variable behavior. 
In the worst case, the behavior of The Guardian's Twitter account was so 
inconsistent that 47 out of 100 messages would have been flagged by \compa as 
malicious. 
We suspect that these accounts are not used by a single person, but instead are managed by a set of
different actors who have different preferences in terms of Twitter clients and
slightly different editing styles.  Our system is currently not able to
characterize accounts with such multi-actor behavior patterns. In the 
general case of a single user operating a given account, however, \compa can 
reliably detect and block changes of behavior.

\subsection{Regular Accounts}
\label{sec:regular_accounts}

To assess the consistency of behavioral profiles for regular accounts, we used
\compa{} to create 64,368 behavioral profiles for randomly selected Twitter users over a
period of 44 days. We used the same threshold selected in Section~\ref{sec:popular_accounts} for this experiment.
To this end, every minute, \compa{} retrieved the latest tweet received from the
Twitter stream and built a behavioral profile for the corresponding account.
2,606 (or 4\%) of these messages violated their account's behavioral profile.
As we would not expect random messages to violate the behavioral profile of the
underlying account, we consider these 4\% the base false discovery rate of
\compa.
Unfortunately, a 4\% false discovery rate is exceedingly high for a practical
deployment of a detection system such as \compa.
Thus, when dealing with regular accounts, instead of detecting compromises of individual user accounts, \compa first
groups accounts by means of message similarity into large-scale {\em campaigns}.
\compa declares members of a campaign as compromised only if a significant
fraction of messages within that campaign violate their respective behavioral
profiles.
%

\noindent{\bf Detecting Large-scale Social Network Compromises} \label{sec:grouping}
\todo[inline]{Show that regular accounts are less consistent than high-profile
ones, and that we need to group messages together. Which is still good, because
this way we can detect large-scale compromises}
A single message that violates the behavioral profile of a user does
not necessarily indicate that this user is compromised and the message
is malicious. The message might merely reflect a normal change of
behavior. For example, a user might be experimenting with new client
software or expanding her topics of interest. Therefore, before we
flag an account as compromised, we require that we can find a number
of similar messages (within a specific time interval) that also
violate the accounts of their respective senders.

Hence, we use message similarity as a second component to distinguish malicious
messages from spurious profile violations. This is based on the assumption that
attackers aim to spread their malicious messages to a larger victim population.
In the following section, we discuss how our system groups together similar messages and assesses their maliciousness.

\section{Detecting Large-scale Social Network Compromises}

\subsection{Grouping Messages}

\megele{No longer mentioned previously.} 
To perform this grouping of messages, we can either first group similar messages
and then check all clustered messages for behavioral profile violations, or we
can first analyze all messages on the social network for profile violations and
then cluster only those that have resulted in violations. The latter approach
offers more flexibility for grouping messages, since we only need to examine the
small(er) set of messages that were found to violate their user profiles. This
would allow us to check if a group of suspicious messages was sent by users that
are all directly connected in the social graph, or whether these messages were
sent by people of a certain demographics. Unfortunately, this approach requires
to check {\it all} messages for profile violations. While this is certainly
feasible for the social networking provider, our access to these sites is
rate-limited in practice.  Hence, we need to follow the first approach: More
precisely, we first group similar messages. Then, we analyze the messages in
clusters for profile violations. To group messages, we use the two simple
similarity measures, discussed in the following paragraphs.

\noindent{\bf Content similarity.} Messages that contain similar text can be considered
related and grouped together. To this end, our first similarity measure uses
n-gram analysis of a message's text to cluster messages with similar contents.
We use entire words as the basis for the n-gram analysis. Based on initial tests
to evaluate the necessary computational resources and the quality of the
results, we decided to use four-grams. That is, two messages are considered
similar if they share at least one four-gram of words (i.e., four consecutive,
identical words). 

\noindent{\bf URL similarity.} This similarity measure considers two messages to be
similar if they both contain at least one link to a similar URL. The na\"ive
approach for this similarity measure would be to consider two messages similar
if they contain an identical URL. However, especially for spam campaigns, it is
common to include identifiers into the query string of a URL (i.e., the part in
a URL after the question mark).
Therefore, this similarity measure discards the query string and relies on the
remaining components of a URL to assess the similarity of messages. Of course,
by discarding the query string, the similarity measure might be incorrectly
considering messages as similar if the target site makes use of the query string
to identify different content. Since {\tt YouTube} and {\tt Facebook} use the
query string to address individual content, this similarity measure discards
URLs that link to these two sites.

Many users on social networking sites use URL shortening services while adding
links to their messages. In principle, different short URLs could point to the
same page, therefore, it would make sense to expand such URLs, and perform the
grouping based on the expanded URLs. Unfortunately, for performance reasons, we
could not expand short URLs in our experiments. On Twitter, we observe several
million URLs per day (most of which are shortened). This exceeds by far the
request  limits imposed by any URL shortening service.

We do not claim that our two similarity measures represent the only ways in which
messages can be grouped. However, as the evaluation in
Section~\ref{sec:evaluation} shows, the similarity measures we chose perform very well in
practice. Furthermore, our system can be easily extended with additional
similarity measures if necessary.

\subsection{Compromised Account Detection}
\label{sec:detection}
\megele{this seems too long}
Our approach groups together similar messages that are generated in a
certain time interval. We call this the 
\emph{observation interval}. For each group, our system checks all
accounts to determine whether each message violates the corresponding
account's behavioral profile. Based on this analysis, our approach has
to make a final decision about whether an account is compromised or
not.

\noindent{\bf Suspicious groups.} 
A group of similar messages is called a {\em suspicious
group} if the fraction of messages
that violates their respective accounts' behavioral profiles exceeds a threshold
$th$.
In our implementation, we decided to use a threshold that is dependent on the size
of the group. The rationale behind this is that, for small groups, there
might not be enough evidence of a campaign being carried out unless a high
number of similar messages violate their underlying behavioral profiles.
In other words, small groups of similar messages could appear coincidentally,
which might lead to false positives if the threshold for small groups is too low.
This is less of a concern for large groups that share a similar message. In
fact,  even the existence of large groups is already somewhat unusual.
This can be taken into consideration by choosing a lower threshold value for
larger groups.
Accordingly, for large groups, it should be sufficient to raise an alert if a smaller percentage of
messages violate their behavioral profiles. Thus, the threshold $th$ is a linear function of the size of the group
$n$ defined as 
$th(n) = \max(0.1, kn + d)$.

Based on
small-scale experiments, we empirically determined that the parameters
$k=-0.005$ and $d=0.82$ work well. The $max$ expression assures that at least ten
percent of the messages in big groups must violate their behavioral profiles to
get the group's users flagged as compromised. Our
experiments show that these threshold values are robust, as small modifications
do not influence the quality of the results.
Whenever there are more than $th$ messages in a group (where each message violates
its profile), \compa{} declares all users in the group as compromised.

\noindent{\bf Bulk applications.} Certain popular applications, such as \texttt{Nike+} or
\texttt{Foursquare}, use templates to send similar messages to their users.
Unfortunately, this can lead to false positives. 
We call these applications bulk applications.  To identify popular bulk
applications that send very similar messages in large amounts, \compa{} needs to
distinguish regular client applications (which do not automatically post using
templates) from bulk applications. To this end, our system analyzes a randomly
selected set of {S} messages for each application, drawn from {\it all} messages
sent by this application. \compa{} then calculates the average pairwise
Levenshtein ratios for these messages. The Levenshtein ratio is a measure of the
similarity between two strings based on the edit distance. The values range
between 0 for unrelated strings and 1 for identical strings. We empirically
determined that the value 0.35 effectively separates regular client applications
from bulk applications.

\compa{} flags all suspicious groups produced by client applications
as compromised. For bulk applications, a further distinction is
necessary, since we only want to discard groups that are due to
{\it popular} bulk applications. Popular bulk applications constantly
recruit new users. Also, these messages are commonly synthetic, and
they often violate the behavioral profiles of new users. For existing
users, on the other hand, past messages from such applications
contribute to their behavioral profiles, and thus, additional messages
do not indicate a change in behavior. If many users made use of the
application in the past, and the messages the application sent were in
line with these users' behavioral profiles, \compa{} considers such an
application as popular. 

To assess an application's popularity, \compa{} calculates the number
of distinct accounts in the social network that made use of that
application before it has sent the first message that violates a
user's behavioral profile.  This number is multiplied by an age factor
(which is the number of seconds between the first message of the
application as observed by \compa{} and the first message that
violated its user's behavioral profile).  The intuition behind this
heuristic is the following: An application that has been used by many
users for a long time should not raise suspicion when a new user
starts using it, even if it posts content that differs from this
user's established behavior.  Manual analysis indicated that bulk
applications that are used to run spam and phishing campaigns over
compromised accounts have a very low popularity score. Thus, \compa{}
considers a bulk application to be popular if its score is above 1
million. We assume that popular bulk applications do not pose a threat
to their users. Consequently, \compa{} flags a suspicious group as
containing compromised accounts only if the group's predominant
application is a non-popular bulk application.

\section{Evaluation}
\label{sec:evaluation}
We implemented our approach in a tool, called \compa{} and evaluated it on 
Twitter and Facebook; we collected tweets in real time from Twitter, while we
ran our Facebook experiments on a large dataset crawled in 2009.

We show that our system is capable of building meaningful behavioral profiles
for individual accounts on both networks.
By comparing new messages against these profiles, it is possible to detect
messages that represent a (possibly malicious) change in the behavior of the
account.
By grouping together accounts that contain similar messages, many of which
violate their corresponding accounts' behavioral profiles, \compa{} is able to
identify groups of compromised accounts that are used to distribute malicious
messages on these social networks.
Additionally, \compa identifies account compromises without a subsequent
grouping step if the underlying behavioral profile is consistent over time.
We continuously ran \compa{} on a stream of 10\% of all public Twitter messages
on a single computer (Intel Xeon X3450, 16 GB ram).
The main limitation was the number of user timelines we could request from
Twitter, due to the enforced rate-limits.
Thus, we are confident that \compa{} can be scaled up to support online social
networks of the size of Twitter with moderate hardware requirements.

We first detail the dataset we used to perform the evaluation of our work.
Subsequently, we discuss a series of real world account compromises against
popular Twitter accounts that \compa could have prevented, and conclude this
section with an evaluation of large-scale compromises that \compa detected on
the Twitter and Facebook social networks.

\subsection{Data Collection}
\label{sec:data_collection}

\noindent{\bf Twitter Dataset}
We obtained elevated access to Twitter's streaming and RESTful API
services.  This allowed us to collect around 10\% of all public tweets
through the streaming API, resulting in roughly 15 million tweets per
day on average. We collected this data continuously starting May 13,
2011 until Aug 12, 2011.
In total, we collected over 1.4 billion tweets from Twitter's stream. The stream
contains live tweets as they are sent to Twitter. We used an
observation interval of one hour. Note that since the stream contains randomly
sampled messages, \compa{} regenerated the behavioral profiles for all involved
users every hour. This was necessary, because due to the 10\% random sampling it was not guaranteed that we
would see the same user multiple times.

To access the historical timeline data for individual accounts, we
rely on the RESTful API services Twitter provides. To this end,
Twitter whitelisted one of our IP addresses, which allowed us to make
up to 20,000 RESTful API calls per hour. A single API call results in
at most 200 tweets. Thus, to retrieve complete timelines that exceed
200 tweets, multiple API requests are needed. Furthermore, Twitter
only provides access to the most recent 3,200 tweets in any user's
timeline. To prevent wasting API calls on long
timelines, we retrieved timeline data for either the most recent three
days, or the user's 400 most recent tweets, whatever resulted in more
tweets.

On average, we received tweets from more than 500,000 distinct users
per hour. Unfortunately, because of the API request limit, we were not
able to generate profiles for all users that we saw in the data
stream. Thus, as discussed in the previous section, we first cluster
messages into groups that are similar. Then, starting from the largest
cluster, we start to check whether the messages violate the behavioral
profiles of their senders. We do this, for increasingly smaller
clusters, until our API limit is exhausted. On average, the created groups
consisted of 30 messages. This process is then repeated for the next
observation period.

\noindent{\bf Facebook Dataset}
Facebook does not provide a convenient way of collecting data.
Therefore, we used a dataset that was crawled
in 2009. We obtained this dataset from an independent research group
that performed the crawling in accordance with the privacy guidelines
at their research institution. Unfortunately, Facebook is actively preventing
researchers from collecting newer datasets from their platform by various means,
including the threat of legal action. 
This dataset was crawled from geographic networks on
Facebook. Geographic networks were used to group together
people that lived in the same area. The default privacy policy for
these networks was to allow anybody in the network to see all the
posts from all other members. Therefore, it was easy, at the time, to
collect millions of messages by creating a small number of profiles
and join one of these geographic networks. For privacy reasons,
geographic networks have been discontinued in late 2009. The dataset
we used contains \totalFacebookStati{} wall posts collected from five
geographic networks (i.e., London, New York, Los Angeles, Monterey
Bay, and Santa Barbara). These wall posts are distributed over almost
two years (Sept. 2007 - July 2009).

\subsection{Detection on Twitter}

The overall results for our Twitter evaluation are presented in
Table~\ref{tbl:results}. Due to space constraints, we will only
discuss the details for the {\em text similarity measure}
here. However, we found considerable overlap in many of the groups
produced by both similarity measures. More precisely, for over 8,200
groups, the two similarity measures (content and URL similarity)
produced overlaps of at least eight messages. \compa{} found, for
example, phishing campaigns that use the same URLs and the same text
in their malicious messages. Therefore, both similarity measures
produced overlapping groups.
 
The {\em text similarity} measure created \totalTextOracleClusters~groups
with messages of similar content. \numTOCgood~groups were
reported as legitimate, while \numTOCbad~groups were reported as compromised. These
\numTOCbad~groups correspond to \numTOCbadAccounts~compromised accounts.
 Interestingly, only \numTOCApps~of \numApps~applications ever produced tweets that got
grouped together. Furthermore, only \numTOCbadApps~of these applications 
contributed to the groups that were identified as compromised.

For each group of similar messages, \compa{} assessed whether the predominant
application in this group was a regular client or a bulk application. Our system
identified \numTOCBulk~groups in the bulk category, of which \numTOCBulkBad~were flagged as compromised.
Moreover, \compa{} identified a total of \numTOCClient~groups that originated from
client applications. Of these, \numTOCClientBad~were flagged as compromised. 

Overall, our system created a total of \totalNumProfiles~behavioral
profiles. \compa{} identified \totalNumViolations~messages that
violate the behavioral profiles of their corresponding
accounts. Finally, \totalNumDeleted~messages were deleted by the
time our system tried to compare these messages to their respective
behavioral profiles (i.e., within an hour).

\begin{table*} {
\begin{center}
\scalebox{0.8}{

\begin{tabular}{|l|r@{ }l|r@{ }l|r@{ }l|r@{ }l|r@{ }l|r@{ }l|}
  \hline 
  {\bf Network \& Similarity Measure} & \multicolumn{4}{|c|}{\textbf{Twitter Text}} & \multicolumn{4}{|c|}{\textbf{Twitter URL}} & \multicolumn{4}{|c|}{\textbf{Facebook Text}}\\
  \hline
  \hline
  & \textbf{Groups} & & \textbf{Accounts} & & \textbf{Groups} & & \textbf{Accounts} & & \textbf{Groups} & & \textbf{Accounts} & \\
  \hline
  \textbf{Total Number} & \totalTextOracleClusters & & & &
\totalLinkOracleClusters & & & & \totalTextOracleClustersFB & & & \\ 
  \hline 
  \textbf{\# Compromised} & \numTOCbad &  & \numTOCbadAccounts &
& \numLOCbad & & \numLOCbadAccounts & & \numTOCbadFB & & \numTOCbadAccountsFB & \\
  \hline
  \hspace{1cm}\textbf{False Positives} & \numTOCbadFPR\% (\numTOCbadFP) & & 
\numTOCbadAccountsFPR\% (\numTOCbadAccountsFP) & & \numLOCbadFPR\%
(\numLOCbadFP) & & \numLOCbadAccountsFPR\% (\numLOCbadAccountsFP) & & \numTOCbadFPRFB{} (\numTOCbadFPFB) & & \numTOCbadAccountsFPRFB{} (\numTOCbadAccountsFPFB) & \\
  \hline
  \textbf{\# Bulk Applications} & \numTOCBulk & & & & \numLOCBulk & & & & N/A & & N/A &\\
  \hline
  \textbf{\# Compromised Bulk Applications} & \numTOCBulkBad & & \numTOCBulkBadAccounts & & 
\numLOCBulkBad & & \numLOCBulkBadAccounts & & N/A & & N/A & \\
  \hline
  \hspace{1cm}\textbf{False Positives} & \numTOCBulkBadFPR\% (\numTOCBulkBadFP) & & 
\numTOCBulkBadAccountsFPR\% (\numTOCBulkBadAccountsFP) & & \numLOCBulkBadFPR\%
(\numLOCBulkBadFP) & & \numLOCBulkBadAccountsFPR\% (\numLOCBulkBadAccountsFP) & & N/A & & N/A & \\
  \hline
  \textbf{\# Client Applications} & \numTOCClient & & & & \numLOCClient & & & & N/A & & N/A & \\
  \hline
  \textbf{\# Compromised Client Applications} & \numTOCClientBad & &
\numTOCClientBadAccounts & & \numLOCClientBad & & \numLOCClientBadAccounts & & N/A & & N/A &\\
  \hline
  \hspace{1cm}\textbf{False Positives} & \numTOCClientBadFPR\% (\numTOCClientBadFP) & &
\numTOCClientBadAccountsFPR\% (\numTOCClientBadAccountsFP) & &
\numLOCClientBadFPR\% (\numLOCClientBadFP) & & \numLOCClientBadAccountsFPR\%
(\numLOCClientBadAccountsFP) & & N/A & & N/A &\\
  \hline

\end{tabular}
}
\end{center}
  \caption{\small Evaluation Results for the Text (Twitter and Facebook) and URL (Twitter) Similarity measure.}
  \label{tbl:results} }
\end{table*} 

\noindent\textbf{False Positives}
\label{sec:accuracy}
Using the text similarity measure, \compa{} identified
\numTOCbadAccounts{} compromised Twitter accounts in \numTOCbad{}
clusters. We performed an exhaustive false positive analysis of \compa in our
previous work~\cite{egele13:compa}. Due to space limitations, we omit repeating
this description here.
In summary, \numTOCbadFP~of the \numTOCbad~groups (\numTOCbadFPR\%) that \compa{}
flagged
as containing compromised are labeled as 
false positives. Note that each group consists of multiple tweets, each from a
different Twitter account. Thus, the above mentioned results are equivalent to
flagging \numTOCbadAccounts~user as compromised, where \numTOCbadAccountsFP~(\numTOCbadAccountsFPR\%) are false positives.

One characteristic that directly affects the probability of a false positive
detection is the length of the message stream that is used to learn the
behavioral profile. Intuitively, the longer a user's messages stream is, the
more comprehensive is the resulting behavioral profile. For a detailed
discussion and analysis of this intuition, we again refer to~\cite{egele13:compa}.

\noindent{\bf False Negatives} \label{sec:falsenegatives} 
Precisely assessing false negatives in large datasets, such as the ones we are
evaluating \compa on, is a challenging endeavor. However, we found after
extensive sampling (64,000 random accounts) that the grouping feature in \compa
did not cause undue amounts of false negatives. In our previous work we detail
our analysis to conclude that \compa suffers from roughly 4\% false negatives in
detecting compromised accounts of regular Twitter users.
\subsection{Detection on Facebook}
As the Facebook dataset spans almost two years we increased the {\em observation
interval} to eight hours to cover this long timespan. Furthermore, we only
evaluated the Facebook dataset with the text similarity measure to group similar
messages.

Our experiments indicated that a small number of popular applications resulted
in a large number of false positives. Therefore, we removed the six most popular
applications, including {\em Mafia Wars} from our dataset. Note that these six
applications resulted in groups spread over the whole dataset. Thus, we think it
is appropriate for a social network administrator to white-list applications at
a rate of roughly three instances per year.

In total, \compa{} generated \totalNumProfilesFB{} profiles in
\totalTextOracleClustersFB{} groups and  flagged \numTOCbadFB{} groups
as compromised (i.e, \numTOCbadAccountsFB{} compromised accounts). All flagged groups
are created by bulk applications. \numTOCbadFPFB{} legitimate groups were
incorrectly classified (i.e., \numTOCbadFPRFB{} false positives) as compromised; they contained \numTOCbadAccountsFPFB{} (\numTOCbadAccountsFPRFB{}) users. 

\subsection{Case studies} 
\label{sec:casestudies} 
As mentioned in Section~\ref{sec:popular_accounts}, \compa successfully
detected four high-profile Twitter compromises. In the following, we discuss
those incidents in more detail, highlighting what type of anomalies were picked
up by \compa compared to the typical behavior of these accounts. In addition, we
discuss a compromise that was simulated by the fast-food company Chipotle on
their Twitter account, for promotional reasons. We demonstrate that in this case
the message did not show particular anomalies compared to the typical behavior
of the account, and therefore \compa would have correctly detected it as being
authored by their legitimate owners.

\noindent\textbf{Associated Press.}
Comparing the malicious message message against the behavioral profile of the @AP account resulted in significant differences among many features that our system
evaluates. For example, the fake news was posted via the Twitter website,
whereas the legitimate owners of the @AP account commonly use the SocialFlow
application to send status updates. Furthermore, the fake tweet did
not include any links to additional information, a practice that the @AP account
follows very consistently. 

Only two features in our behavioral model did not signify a change of behavior.
The time when the tweet was sent (i.e., 10:07UTC) and the language of the tweet
itself. The authors of the @AP account as well as the attackers used the English
language to author their content. While the language is undoubtedly the same, a
more precise language analysis could have determined an error in capitalization
in the attacker's message. 

\noindent\textbf{FoxNews Politics.} 
This tweet violated almost all the features used by our system. For example, the
tweet was sent in the middle of the night (i.e., 23:24UTC), through the main
Twitter web site. Furthermore, it did not include a link to the full story on
the Fox News website. The tweet also made extensive use of hashtags and
mentions, a practice not commonly used by the @foxnewspolitics account.

\noindent\textbf{Skype.}
\compa successfully detected the compromise because the offending message
significantly diverged from the behavioral profile constructed for the Skype
account.
The only two features that did not diverge from the behavioral profile were the
time and language information.
Since the Skype profile as well as the malicious message were authored in
English, \compa did not detect a deviation in this feature.
More interestingly, however, the time the message was sent, perfectly aligned
with the normal activity of the Skype account.
We would assume that an observant legitimate owner of the account would
detect such a malicious message during their regular activity.
However, presumably because of the holiday season, it took more than two hours
before the offending message was removed by the legitimate owners of the Skype
account.
In the meantime, the offending message got retweeted over 8,000 times.
This incident prominently demonstrates the advantages an automated technique for
the detection of compromised accounts would entail, as such attacks can have
significant negative impact on a brand's online reputation. 

\noindent\textbf{Yahoo! News.} 
Our system detected significant deviations of the offending message when
compared to the extracted behavioral profile for the account. 
Similarly to the above mentioned cases, the attackers used Twitter's web portal
to send the offending messages, whereas YahooNews predominantly relies on the
TweetDeck application to post new content.
While YahooNews frequently links to detailed information and often mentions
their source by using the direct communication feature (i.e., @-mentions), the
offending tweets featured neither of these characteristics.

\noindent\textbf{Chipotle.}
On July 21, 2013 multiple news websites reported that the main Twitter account
of the Chipotle Mexican Grill restaurant chain got compromised\footnote{At the
time of writing, the public timeline of the
\href{https://twitter.com/chipotletweets}{@chipotletweets} account still
contains these tweets.}.
Indeed, twelve ``unusual'' successive messages were posted to the
@chipotletweets account that day, before the apparent legitimate operator
acknowledged that they experienced issues with their account.
Because this was in the midst of other compromises of high-profile accounts
(e.g., Jeep, Burger King, and Donald Trump), this alert seemed credible.
However, when we ran \compa on these twelve messages in question, only minor
differences to the behavioral profile of the Chipotle account emerged.
More precisely, the offending messages  did not contain any direct user
interaction (i.e., mentions) -- a feature prominently used by the legitimate
operator of that account.
However, because this was the only difference compared to the learned behavioral
profile, \compa's classifier did not consider the deviation significant enough
to raise a warning about an account compromise.
Interestingly, three days later, Chipotle acknowledged that they had faked the
account compromise as a publicity measure~\cite{chipotle:fake}.
This illustrates that even trying to fake an account compromise is a non-trivial
endeavor. 
As mentioned, all other features besides the direct user interaction were
perfectly in line with the behavioral profile.
When we investigated the application source model for the Chipotle account we
learned that it is almost exclusively managed via the SocialEngage client
application.
Thus, for an attacker to stealthily compromise Chipotle's account, he would also
have to compromise Chipotle's SocialEngage account.
A similar attempt of faking an account compromise staged by
MTV~\cite{mtv:fake_compromise} did also not result in \compa raising an alert.
Because of our limited view of Twitter's traffic (i.e., we only see a random
10\% sample), we could not evaluate the faked compromise of the BET account
staged in the same campaign by the same actors.

\section{Limitations}
\label{sec:limitations}
An attacker who is aware of \compa{} has several possibilities to prevent
his compromised accounts from being detected by \compa{}. First, the attacker can
post messages that align with the behavioral profiles of the compromised accounts.
As described in Section~\ref{sec:training}, this would require the attacker to
invest significant time and computational resources to gather the necessary
profile information from his victims. Furthermore, social networks have mechanisms in
place that prevent automated crawling, thus slowing down such data gathering
endeavors.

In the case of \compa protecting regular accounts
an attacker could send messages that evade our similarity measures, and thus, although
such messages might violate their compromised accounts' behavioral profiles, they
would not get grouped together. To counter such evasion attempts, \compa{} can
be easily extended with additional and more comprehensive similarity measures. For example, it would be
straight-forward to create a similarity measure that uses the landing page instead of the
URLs contained in the messages to find groups of similar messages. Furthermore,
more computationally expensive similarity measures, such as text shingling or
edit distances for text similarity can also be implemented. Other
similarity measures might leverage the way in which messages propagate along the
social graph to evaluate message similarity. 

\section{Related Work}
\label{sec:related}

The popularity of social networks inspired many scientific studies in both,
networking and security.
Wilson et al. ran a large-scale study of Facebook users~\cite{Wilson:09:eurosys}, while Krishnamurthy et al. provide a characterization
of Twitter users~\cite{Krishnamurthy:2008:usenix}. Kwak et al. analyze the
differences between Twitter and the more traditional social networks~\cite{Kwak:2010:10:www}.

Yardi et al.~\cite{twitterspam} ran an experiment on the propagation of spam on Twitter. 
Their goal was to study how spammers use popular topics in their messages to
reach more victims.
To do this, they
created a hashtag and made it trending, and observed that spammers started using
the hashtag in their messages.

Early detection systems for malicious activity on social networks
focused on identifying fake accounts and spam
messages~\cite{Benvenuto:2010:CEAS,Lee:2010:SIGIR,Stringhini:10:socialnet-spam} 
by leveraging features that are geared towards recognizing characteristics
of spam accounts (e.g., the presence of URLs in messages or message similarity
in user posts). Cai et
al.~\cite{Cai:2012:ndss} proposed a system that detects fake profiles
on social networks by examining densely interconnected groups of profiles. These techniques work reasonably well, and both
Twitter and Facebook rely on similar heuristics to detect fake
accounts ~\cite{facebookrules,twitterrules}.

In response to defense efforts by social network providers, the focus
of the attackers has shifted, and a majority of the accounts carrying
out malicious activities were not created for this purpose, but
started as legitimate accounts that were
compromised~\cite{Gao:2010:imc,Grier:2010:ccs}. Since these accounts
do not show a consistent behavior, previous systems will fail to
recognize them as malicious. Grier et al.~\cite{Grier:2010:ccs}
studied the behavior of compromised accounts on Twitter by entering
the credentials of an account they controlled on a phishing campaign
site. 
This approach does not scale as it requires identifying and joining each new phishing
campaign. 
Also, this approach is limited to phishing campaigns.  Gao et
al.~\cite{Gao:2010:imc} developed a clustering approach to detect spam
wall posts on Facebook. They also attempted to determine whether an
account that sent a spam post was compromised. To this end, the
authors look at the wall post history of spam accounts.  However, the
classification is very simple. When an account received a benign wall
post from one of their connections (friends), they automatically
considered that account as being legitimate but compromised. The
problem with this technique is that previous work showed that spam
victims occasionally send messages to these spam
accounts~\cite{Stringhini:10:socialnet-spam}. This would cause their
approach to detect legitimate accounts as compromised.  Moreover, the
system needs to know whether an account has sent spam before it can
classify it as fake or compromised. Our system, on the other hand,
detects compromised accounts also when they are not involved in spam
campaigns.  As an improvement to these techniques, Gao et
al.~\cite{Gao:2012:ndss} proposed a system that groups similar
messages posted on social networks together, and makes a decision
about the maliciousness of the messages based on features of the
message cluster. Although this system can detect compromised accounts,
as well as fake ones, their approach is focused on detecting accounts
that spread URLs through their messages, and, therefore, is not as
generic as \compa{}.

Thomas et al.~\cite{Thomas:2011:oakland} built Monarch to detect malicious messages on social
networks based on URLs that link to malicious sites.
By relying only on URLs, Monarch 
misses other types of malicious messages. For example, our previous
work~\cite{egele13:compa}
illustrates that \compa detects scams based on phone numbers and 
XSS worms spreading without linking to a malicious URL. 

\textsc{WarningBird}~\cite{Lee:2012:ndss} is a system that 
detects spam links posted on Twitter by analyzing the characteristics of
HTTP redirection chains that lead to a final spam page.

Xu et al.~\cite{Xu:2010:acsac} present a system that, by monitoring a small
number of nodes, detects worms propagating on social networks. 
This paper does not directly address the problem
of compromised accounts, but could detect large-scale
infections such as \emph{koobface}~\cite{koobface}.
Chu et al.~\cite{Chu:2010:acsac} analyze three
categories of Twitter users: humans, bots, and cyborgs, which are software-aided
humans that share characteristics from both bots and humans.
To this end, the authors use a classifier that examines how
regularly an account tweets, as well as other account features such as the
application that is used to post updates. Using this paper's terminology,
compromised accounts
would fall in the cyborg category. However, the paper does not provide a way of
reliably detecting them, since these accounts are often times misclassified as
either bots or humans. More precisely, their true positive ratio for
\emph{cyborg} accounts is only of 82.8\%. In this paper, we showed that we can
detect such accounts much more reliably. 
Also, the authors in~\cite{Chu:2010:acsac} do not provide a clear
distinction between compromised accounts and legitimate ones that use
third-party applications to post updates on Twitter.

Yang et al.~\cite{Yang:11:Die} studied new Twitter spammers that act in a
stealthy way to avoid detection. In their system, they use advanced features
such as the topology of the network that surrounds the spammer. They do not try
to distinguish compromised from spam accounts. 

Recent work in the online abuse area focused on detecting accounts that are
accessed by botnets, by either looking at accounts that are accessed by many
IP addresses~\cite{stringhini2015evilcohort} or by looking at accounts that
present strong synchronized activity~\cite{cao2014uncovering}. \compa can
detect compromised accounts that are accessed by botnets as well, but has the
additional advantage of being able to identify and block hijacked accounts
that are used in isolation.

\section{Conclusions}

In this paper, we presented \compa, a system to detect compromised
accounts on social networks.  \compa uses statistical
models to characterize the behavior of social network users,
and leverages anomaly detection techniques to identify sudden changes in their
behavior. The
results show that our approach can reliably detect compromises affecting high-
profile social network accounts, and can detect compromises of regular 
accounts, whose behavior is typically more variable, by aggregating together 
similar malicious messages.
\Urlmuskip=0mu plus 1mu\relax
\bibliographystyle{IEEEtran} 
\bibliography{biblio} 
\end{document}